\documentclass[traditabstract]{aa}  
\usepackage{graphicx}
\usepackage{txfonts}
\usepackage{rotating}
\usepackage{natbib,twoopt}
\usepackage{breqn}

\usepackage{subfig}

\begin{document}

\title{A Lucky Imaging search for stellar companions to transiting planet host stars}

   \author{ Maria W\"ollert\inst{1} \and
            Wolfgang Brandner\inst{1} \and
            Carolina Bergfors \inst{2} \and
            Thomas Henning\inst{1}
           }

   \institute{ Max-Planck-Institut f\"ur Astronomie, K\"onigstuhl 17,
               69117 Heidelberg, Germany\\
               \email{woellert@mpia.de} 
               \and Department of Physics and Astronomy, University College London, Gower Street, London WC1E 6BT, United Kingdom
              }

   \date{Received ---; accepted ---}

 
   \abstract{The presence of stellar companions around planet hosting stars influences the architecture of their planetary systems. To find and characterise these companions and determine their orbits is thus an important consideration to understand planet formation and evolution. For transiting systems even unbound field stars are of interest if they are within the photometric aperture of the light curve measurement. Then they contribute a constant flux offset to the transit light curve and bias the derivation of the stellar and planetary parameters if their existence is unknown. Close stellar sources are, however, easily overlooked by common planet surveys due to their limited spatial resolution. We therefore performed high angular resolution imaging of $49$ transiting exoplanet hosts to identify unresolved binaries, characterize their spectral type, and determine their separation. The observations were carried out with the Calar Alto $2.2\,$m telescope using the Lucky Imaging camera AstraLux Norte. All targets were imaged in $i^{\prime}$ and $z^{\prime}$ passbands. We found new companion candidates to WASP-14 and WASP-58, and we re-observed the stellar companion candidates to CoRoT-2, CoRoT-3, CoRoT-11, HAT-P-7, HAT-P-8, HAT-P-41, KIC 10905746, TrES-2, TrES-4, and WASP-2. We deduce from the stellar density around all sources that $two$ companion candidates out of the targets with the first position measurement (CoRoT-11, HAT-P-41, KIC 10905746, WASP-14 and WASP-58) are probably unbound.
  In addition, we re-analyse the influence of the sources close to WASP-14 and WASP-58 on the planetary parameters given in the literature and find no significant changes.

  }

   \titlerunning{A Lucky Imaging search for stellar companions to TEP host stars}
   \maketitle
%

\section{Introduction}

Since the first detection of a transiting planet around a main-sequence star in 2000 \citep{Henry00}, many hundreds of giant planets have been found by transit surveys. Interestingly, a vast number of these planets are located close to their host star with semi-major axes smaller than $0.5\,$AU. In addition, many of them are not in circular, well aligned orbits, but show a variety of eccentricities and orbit inclinations. In fact, about $30 \%$ of the investigated systems have been found to be on orbits that are misaligned with respect to the stellar equator \citep[see][for a list of investigated systems]{Southworth11}. 

Giant planets are, however, believed to form beyond the ice line in a well-aligned circumstellar disk at several AU from their host star. This means that they have to move inwards at a later stage. Several mechanisms have been discussed  to explain the findings. First of all, migration, which is the interaction of the embedded planet with the surrounding disk \citep[see][for a review]{Baruteau13}, can cause inward-spiraling of the planet. Yet, disk-planet interaction in a planetary system with only one massive planet tends to damp the planet's eccentricity and inclination and cannot account for eccentric or inclined close-in orbits. Planet-planet interaction in multiple planetary systems can, however, excite the eccentricities and inclinations of the planetary orbits as has been shown in numerical simulations \citep[e.g.][]{Ford08}. Similar but stronger effects are caused by a companion star that may interact with the planets around the primary via the Lidov-Kozai mechanism \citep{Naoz11}. It has been shown in simulations that both tight and wide binary companions can strongly perturb the formation and evolution of planetary systems  \citep{Bouwman06, Fabrycky07, Malmberg11, Petrovich14}. In addition, the tidal field of the Milky Way induces a pseudo-random walk of the pericenters of wide binaries leading to strong perturbations of the planets in the system \citep{Kaib13}. It is thus crucial to look for close and wide stellar companions and correlate their properties with those of the close-in giant planets to understand the influence of binarity on the evolution of planetary systems. A fair number of transiting exoplanet host stars (TEPs) have already been subject to high angular resolution studies \cite[e.g.][]{Daemgen09, Lillo-Box12, Adams12, Bergfors13, Faedi13, Ginski13, Law13} and a couple of gravitationally bound companions has been found. The increasing number of transiting exoplanet detections enables, however, a more thorough analysis of planetary systems in binaries. 

Another important aspect of the search for stellar companions is the dilution of the light curve by a blended companion, no matter if it is bound or not. An unresolved faint secondary star within the photometric aperture dilutes the measured transit or occultation depth by the fraction $F_{\textrm{\tiny{bright}}} / F_{\textrm{\tiny{total}}}$. For example, \cite{Crossfield12} showed that an M dwarf near WASP-12 diluted the planet's infrared emission spectrum by up to 15\%, an effect substantially larger than the previously quoted uncertainty on the planet's emission. \cite{Daemgen09, Narita10, Bergfors13} have reported optical imaging of stars located near several transit systems whose hot Jupiters have well-characterized spectra: HAT-P-7, TrES-2, TrES-4, WASP-2, and WASP-12.

In this paper, we present the results of our photometric and astrometric Lucky Imaging study of another $49$ transiting exoplanet host stars, which complements the study of \cite{Daemgen09} and \cite{Bergfors13}. Besides the derivation of $i$ and $z$ magnitudes, photometric spectral types, and the relative position of the companion candidates, we discuss the probability of physical companionship of the observed sources and the effect of the flux contribution a companion candidate has on stellar and planetary parameters.

\section{Observations and data reduction}

\subsection{Lucky Imaging with AstraLux}

The $49$ transiting exoplanet host stars were observed at the $2.2\,$m telescope at Calar Alto in June $24-28$, 2013. We imaged each target  with the Lucky Imaging camera AstraLux Norte \citep{Hormuth08} in SDSS $i^{\prime}$ and $z^{\prime}$-passband. Lucky Imaging enables nearly diffraction-limited observations by taking a series of many short exposures of the target and combining only the least distorted frames to the final image. Depending on the target brightness, we obtained between $20 \, 000$ and $60 \, 000$ individual frames with exposure times of $15\,$ms each. The $10\%$ of images with the highest Strehl ratio were then shifted and added to produce the final image. The field of view of the final image was about $12''$x$12''$ with the transiting exoplanet system at the centre. As indicated by the {\it 1st epoch} flag in Table \ref{no_comp}, we observed 42 of the 49 stars  with Lucky Imaging for the first time.

Astrometric reference stars in the globular cluster M$\,13$ were observed at the beginning and the middle of each night to calibrate the detector rotation and pixel scale. The plate scale and detector rotation was deduced from the measured detector position of five stars of M$\,13$ with the image reduction and analysis facility IRAF using the task imexamine. We calculated the separation and rotation angle of the five stars pairwise and compared these values to data from high-quality astrometric HST/ACS observations. This analysis yielded a plate scale of $23.48 \pm 0.01$ mas/px and a detector rotation of $1.7^{\circ} \pm 0.1^{\circ}$ west of north. Additionally, twilight sky-flats and bias were obtained in both filters at the beginning of each observing night. 

\subsection{Photometry and astrometry}

\begin{figure}[htb]
    \begin{center}
        \includegraphics[width=0.5\textwidth]{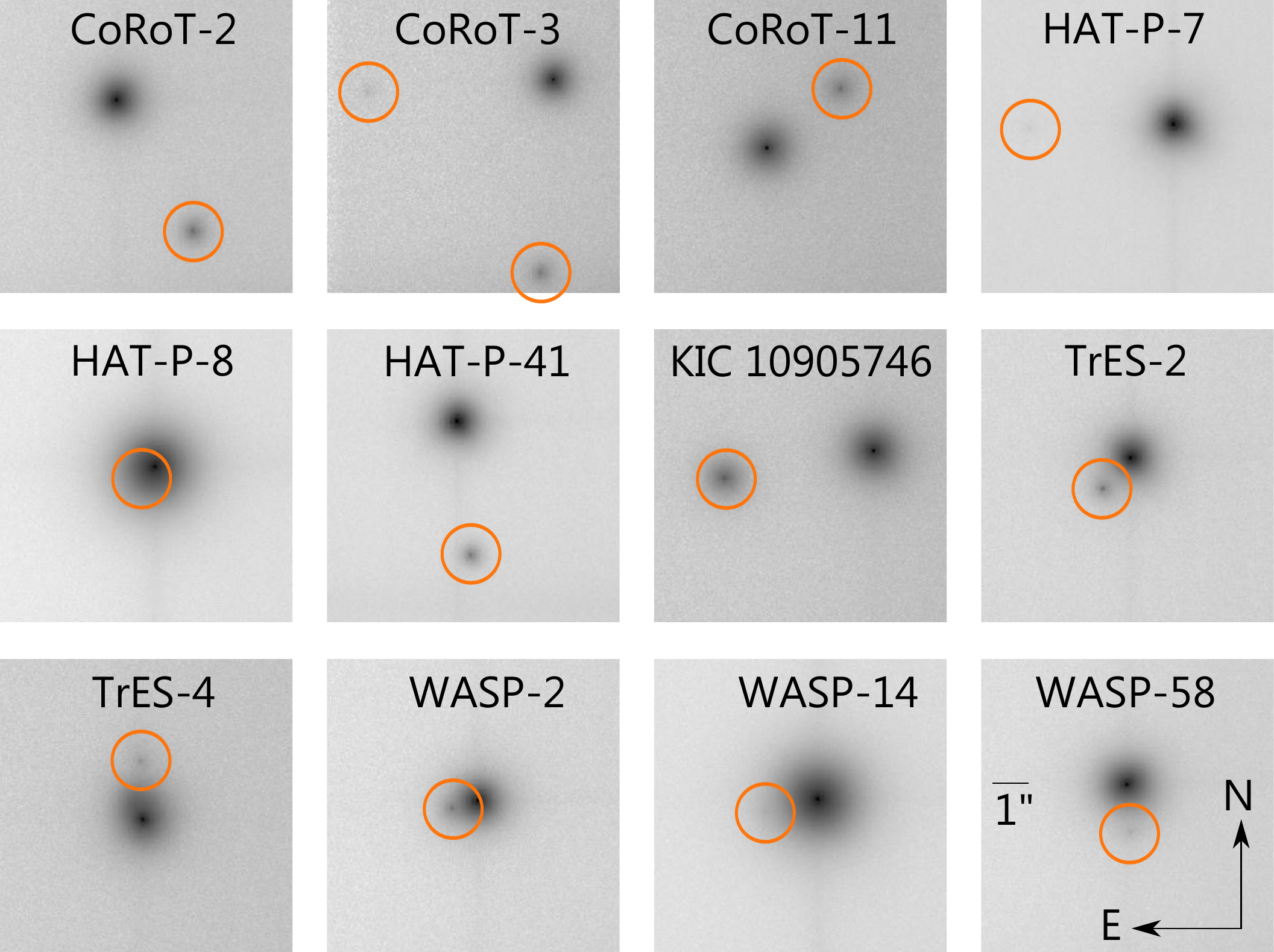}
    \end{center}
    \caption[]{The $z^{\prime}$ filter images of the $12$ exoplanet host stars for which companion candidates have been detected. The image scale is logarithmic. The size and orientation is identical for all images and indicated in the lower right image.} 
    \label{z_images}
\end{figure}

In our sample of $49$ transiting exoplanet host stars we found $13$ companion candidates by visual inspection of the reduced Lucky Imaging frames; they are listed in Table~\ref{with_comp} and the $z^{\prime}$-band frames are depicted in Figure~\ref{z_images}. Eleven of them were previously known: CoRoT-2 \citep{Alonso08}, CoRoT-3 \citep{Deleuil08}, CoRoT-11 \citep{Gandolfi10}, HAT-P-7 \citep{Narita10}, HAT-P-8 \citep{Bergfors13}, HAT-P-41 \citep{Hartman12}, KIC 10905746 \citep{Fischer12}, TrES-2 \citep{Daemgen09}, TrES-4 \citep{Daemgen09}, and WASP-2 \citep{Collier07}. The candidate companions to WASP-14 and WASP-58 have not been reported in the literature so far.

We obtained the positions of the two stellar components by fitting a model psf to the data. The theoretical psf is built from the ideal point spread function without atmospheric turbulences and telescope aberrations, which is then convolved with a Gaussian blurring function. Finally, a Moffat profile is added to the blurred theoretical profile \citep[see][for more information on the procedure]{Woellert14}. The separation and position angle given in Table~\ref{with_comp} is the average value from our $i^{\prime}$ and $z^{\prime}$ measurements; the uncertainties are determined to take the position uncertainty in every single image into account, as well as the uncertainties in pixel scale and instrument rotation. For the very faint companions to HAT-P-8 and WASP-14 the algorithm did not converge and we estimated the position by eye. The companion of HAT-P-8 is itself a binary \citep{Ginski13, Bechter14}, which causes an elongation of the psf and  might be the reason for the non-convergence. 

\begin{table}[htb]
    \caption[]{All transiting exoplanet systems investigated, radial detection limits, and y/n flag indicating  only if 1st epoch observations are available.}
    \label{no_comp}
      \begin{center}
        \begin{tabular}{l l l l l c}
          \hline
          \hline
                        &  \multicolumn{4}{c}{$5 \sigma$ detection limit ($\Delta z^{\prime}$ [mag])} & 1st \\
        Name            & $0.25''$      & $0.5''$        & $1.0''$      & $2.0''$ & epoch \\
        \hline
        \multicolumn{5}{l}{Targets without candidate companions} \\
        \hline
        HAT-P-3 &      3.84     &      4.62     &      5.55     &      6.20 & y\\   
        HAT-P-4 &      3.97     &      4.74     &      5.56     &      6.19 & y\\   
        HAT-P-12        &      3.69     &      4.45     &      5.22     &      5.65 & y\\ 
        HAT-P-14        &      3.99     &      5.04     &      6.47     &      7.32 & y\\ 
        HAT-P-16        &      4.05     &      4.88     &      5.87     &      6.63 & y\\ 
        HAT-P-17        &      3.84     &      4.95     &      6.29     &      7.09 & y\\ 
        HAT-P-18        &      3.64     &      4.43     &      5.24     &      5.56 & y\\ 
        HAT-P-21        &      3.75     &      4.35     &      5.18     &      5.72 & y\\ 
        HAT-P-22        &      3.95     &      4.97     &      6.12     &      7.19 & y\\ 
        HAT-P-23        &      3.50     &      4.19     &      5.06     &      5.43 & y\\ 
        HAT-P-26        &      3.64     &      4.59     &      5.62     &      6.15 & y\\ 
        HAT-P-27&      3.73     &      4.35     &      5.20     & 5.63 & y\\     
        HAT-P-31        &      3.98     &      4.75     &      5.68     &      6.16 & y\\ 
        HAT-P-32        &      3.17     &      3.91     &      4.31     &      4.48 & y\\ 
        HAT-P-34        &      4.05     &      5.40     &      6.80     &      7.38 & y\\ 
        HAT-P-36        &      3.63     &      4.53     &      5.27     &      5.64 & y\\ 
        HAT-P-40        &      3.72     &      4.45     &      5.21     &      5.89 & y\\ 
        HD 97658        &      3.43     &      5.03     &      6.06     &      7.36 & y\\ 
        Kepler-8        &      3.19     &      3.88     &      4.35     &      4.49 & y\\ 
        Kepler-17       &      3.29     &      3.49     &      4.29     &      4.40 & y\\ 
        Kepler-25       &      4.22     &      5.30     &      6.55     &      7.11 & y\\ 
        Kepler-50       &      3.95     &      5.49     &      6.70     &      7.23 & y\\ 
        Qatar-1 &      3.65     &      4.43     &      5.27     &      5.55 & y\\   
        WASP-10 &      3.62     &      4.29     &      5.04     &      5.49 & y\\   
        WASP-21 &      3.67     &      4.52     &      5.34     &      5.90 & y\\   
        WASP-24 &      3.86     &      4.65     &      5.56     &      6.06 & y\\   
        WASP-28 &      3.20     &      3.77     &      4.43     &      4.96 & y\\   
        WASP-37 &      3.13     &      3.65     &      4.20     &      4.69 & y\\   
        WASP-38 &      3.83     &      5.27     &      6.96     &      7.90 & y\\   
        WASP-39 &      3.42     &      4.27     &      5.08     &      5.55 & y\\   
        WASP-47 &      3.71     &      4.46     &      5.36     &      5.71 & y\\   
        WASP-48 &      3.94     &      5.02     &      5.84     &      6.09 & y\\   
        WASP-52 &      3.91     &      4.54     &      5.39     &      5.85 & y\\   
        WASP-59 &      3.62     &      4.34     &      5.27     &      5.68 & y\\   
        WASP-60 &      3.47     &      4.20     &      5.04     &      5.42 & y\\   
        WASP-80 &      4.01     &      5.03     &      6.11     &      6.64 & y\\   
        XO-1    &      3.97     &      4.85     &      5.79     &      6.46 & y\\   
        \hline
        \multicolumn{5}{l}{Targets with candidate companions} \\
        \hline
        CoRoT-2 &      4.10     &      4.83     &      5.73     &      5.99 & n\\   
        CoRoT-3 &      3.69     &      4.41     &      5.29     &      5.81 & n\\   
        CoRoT-11        &      3.29     &      4.17     &      4.82     &      5.01 & y\\ 
        HAT-P-7 &      4.03     &      5.51     &      6.86     &      7.41 & n\\   
        HAT-P-8 &      4.24     &      5.13     &      6.26     &      7.11 & n\\   
        HAT-P-41        &      4.00     &      5.48     &      6.74     &      7.20 & y\\ 
        KIC-10905746    &      3.70     &      4.46     &      5.26     &      5.56 & y\\
        TrES-2  &      4.03     &      5.06     &      6.14     &      6.58 & n\\   
        TrES-4  &      3.62     &      4.40     &      5.32     &      5.78 & n\\   
        WASP-2  &      3.93     &      4.97     &      6.00     &      6.40 & n\\   
        WASP-14 &      4.20     &      4.99     &      6.14     &      7.07 & y\\   
        WASP-58 &      4.36     &      5.30     &      6.19     &      6.60 & y\\
          \hline
        \end{tabular} 
      \begin{quote}
          HAT-P-17 and HAT-P-32 have been observed as well by \cite{Adams13} using adaptive optics at the MMT. In agreement with our results they do not see a companion candidate around HAT-P-17, but they find one around HAT-P-32 at $2.87\arcsec$, $PA=109^\circ$ and $\Delta K = 3.4\,$mag.
      \end{quote}
      \end{center}
\end{table}

\begin{table*}[htb]
    \caption[]{Candidate binary systems and their properties} 
    \label{with_comp}
      \begin{center}
        \begin{tabular}{l l l l l l}
          \hline
          \hline
        Name            &  Sep [$\arcsec$]          & PA [$^{\circ}$] & $\Delta i^{\prime}$ [mag]  &   $\Delta z^{\prime}$ [mag]  &  Date of obs. \\
          \hline
        CoRoT-2         & $4.090 \pm 0.005$ & $208.6 \pm 0.2$ & $3.38 \pm 0.05$ & $3.00 \pm 0.05$ & 26 Jun 2013\\
        CoRoT-3-South   & $5.209 \pm 0.006$ & $174.7 \pm 0.2$ & $3.24 \pm 0.06$ & $3.10 \pm 0.06$ & 26 Jun 2013\\
        CoRoT-3-East    & $4.995 \pm 0.156$ & $92.2  \pm 1.6$ & $5.53 \pm 0.13$ & $5.38 \pm 0.17$ & 26 Jun 2013\\
        CoRoT-11        & $2.545 \pm 0.006$ & $307.2 \pm 0.2$ & $2.34 \pm 0.05$ & $2.68 \pm 0.06$ & 26 Jun 2013\\
        HAT-P-7         & $3.843 \pm 0.034$ & $89.8  \pm 0.4$ & $7.58 \pm 0.17$ & $6.88 \pm 0.20$ & 26 Jun 2013\\
        HAT-P-8         & $1.034 \pm 0.067$ & $139.2 \pm 1.2$ & $7.34 \pm 0.10$ [1]& $6.68 \pm 0.07$ [1] & 27 Jun 2013\\
        HAT-P-41        & $3.619 \pm 0.005$ & $184.1 \pm 0.2$ & $3.65 \pm 0.05$ & $3.40 \pm 0.05$ & 26 Jun 2013\\
        KIC 10905746    & $4.075 \pm 0.006$ & $98.7  \pm 0.2$ & $2.42 \pm 0.1$ & $2.31 \pm 0.1$ & 28 Jun 2013\\
        TrES-2          & $1.106 \pm 0.005$ & $136.3 \pm 0.2$ & $3.83 \pm 0.04$ & $3.37 \pm 0.05$ & 25 Jun 2013\\
        TrES-4          & $1.596 \pm 0.017$ & $0.0   \pm 0.6$ & $4.49 \pm 0.08$ & $4.16 \pm 0.09$ & 25 Jun 2013\\
        WASP-2          & $0.711 \pm 0.019$ & $105.1 \pm 0.5$ & $3.51 \pm 0.04$ & $3.26 \pm 0.06$ & 25 Jun 2013\\
        WASP-14         & $1.423 \pm 0.034$ & $102.2 \pm 1.0$ & $5.35 \pm 0.11$ & $5.45 \pm 0.16$ & 26 Jun 2013\\
        WASP-58         & $1.275 \pm 0.015$ & $183.2 \pm 0.4$ & $5.64 \pm 0.18$ & $5.09 \pm 0.13$ & 25 Jun 2013\\
          \hline
        \end{tabular} 
  \begin{quote}
    [1]: \cite{Bergfors13}
      \end{quote}
      \end{center}
\end{table*}

\begin{figure}[htb]
    \begin{center}
        \includegraphics[trim=3cm 4cm 3cm 5cm, clip=true, width=0.5\textwidth]{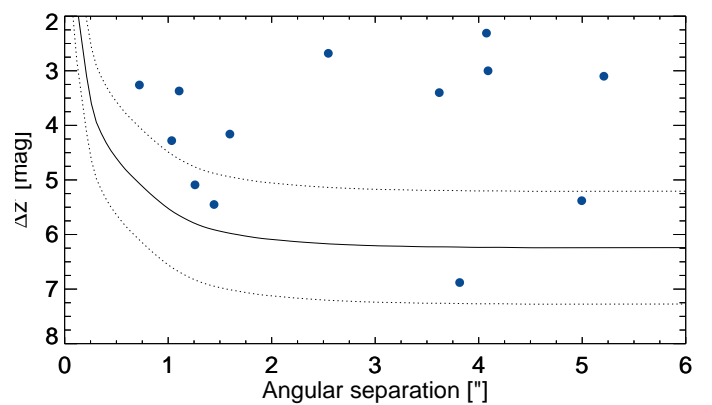}
    \end{center}
    \caption[average $z^{\prime}$ contrast]{Average sensitivity of the survey (solid line). The dashed lines corresponds to $1\sigma$ deviations of the contrast curves. The blue circles indicate the detections.} 
    \label{z_contrast}
\end{figure}

We used the psf-subtracted images to calculate the $5 \sigma$-contrast curve for every target with or without companions; a list of all observed targets and their sensitivity limits at $r=0.25'', 0.5'', 1.0''$ , and $2.0''$ can be found in Table~\ref{no_comp}, the average $z^{\prime}$-passband sensitivity can be found in Fig.~\ref{z_contrast}. The noise of each pixel was estimated using the standard deviation of the residual signal in a box of $5 \times 5\,$pixel around it. This value was then compared to the flux of the star in a $5 \times 5\,$pixel box centred on the peak. The contrast at a specific angular separation range was calculated as the median of the contrast of all pixels in the corresponding distance. 

The brightness difference of the components in the $i^{\prime}$ and $z^{\prime}$ passbands was determined by aperture photometry using the IDL routine APER. We chose an aperture radius of $4.5\,$pixel centred on the stars to ensure that most of the flux is covered. The sky background was measured in an annulus from $6\,$pixel to $8\,$pixel. The error bars in magnitude differences are propagated from the photometric errors estimated with APER.

\section{Companion properties}

\subsection{Spectral types}

The spectral types of the companion candidates are derived from the observed magnitude differences in SDSS $i^{\prime}$ and $z^{\prime}$-band between the host and the companion, and by assuming negligible foreground extinction. The latter is justified as the TEP are located off the Galactic plane and not in the direction of any nearby molecular cloud. The $i^{\prime}$ and $z^{\prime}$ magnitudes of the host stars in turn are estimated from the spectral types given in the literature or, if not known, from 2MASS {\it JHK}-photometry \citep{Cutri03} (HAT-P-41) or from the {\it V-J} colour (KIC-10905746). To convert the spectral type to $i^{\prime}$ and $z^{\prime}$ brightness, we use the {\it ugriz} stellar spectral energy distribution models for main-sequence stars from \cite{Kraus07}. We assume an uncertainty of $0.10\,$mag in both filters to consider the flux uncertainty in $i^{\prime}$ and $z^{\prime}$ due to our estimate. We neglect the small magnitude differences, which arise from the conversion from {\it i,z} to $i^{\prime}$,$z^{\prime}$; they are of the order $0.01\,$mag in both filters for the spectral types of our host stars. Knowing the primaries $i^{\prime}$ and $z^{\prime}$ colours we calculate the $i^{\prime}$ and $z^{\prime}$ magnitude of the companion candidate. For the companion candidates we do not neglect the differences in the photometric system because of their redder colour and we use the conversion formulas on the SDSS webpage to get $i$ and $z$ magnitudes. Since the $r^{\prime}$ magnitude is needed to convert $i^{\prime}$ to $i$, we use a rough estimate of $r^{\prime}$ based on the measured $i^{\prime}$-$z^{\prime}$ colour. This is sufficient since the two photometric systems only deviate strongly for very late sources that we did not detect. 

Assuming a physical association between TEP and companion candidate, we compute absolute $i$ and $z$ magnitudes of the secondary. With the exception of the binary companion candidate to CoRoT-3-east, the absolute magnitudes and the spectral types based on the $i-z$ colour are consistent within the uncertainties (Table~\ref{SpT}). Our spectral type estimates also agree with the values in the literature that are given for some of the targets. The spectral types of the other companion candidates to HAT-P-41, KIC-10905746, and WASP-58 are most likely late-type as expected for a fainter associated companion; for CoRoT-11 and WASP-14 the photometric spectral typing points at early-type stars not at the same distance as the TEP. This makes physical companionship unlikely, but further measurements are needed to give a final answer.

\begin{table*}[htb]
    \caption[]{Spectral type estimates of the candidate companions from $(i-z)$ colour} 
    \label{SpT}
      \begin{center}
        \begin{tabular}{l c c c c c c c l}

          \hline
          \hline
          Name          & SpT(A)& $M_i(A)$ & $M_z(A)$ & $M_i(B)^*$      & $M_z(B)^*$      & $M_i(B)-M_z(B)$       & SpT(B)        & Lit.          \\
          \hline
          CoRoT-2               & G7V   & $4.89 \pm 0.10$ & $5.08 \pm 0.10$ & $8.28 \pm 0.11$       & $8.08 \pm 0.11$       & $0.20 \pm 0.16$       & G4-M0   & M0 [1]                \\
          CoRoT-3-South & F3V   & $3.27 \pm 0.10$ & $3.33 \pm 0.10$ & $6.50 \pm 0.16$       & $6.44 \pm 0.20$       & $0.07 \pm 0.26$       & A2-K7 & K4-K5 [1]               \\
          CoRoT-3-East  & F3V   & $3.26 \pm 0.10$ & $3.33 \pm 0.10$ & $8.78 \pm 0.12$       & $8.72 \pm 0.12$       & $0.07 \pm 0.16$       & F1-K5 &                         \\
          CoRoT-11      & F6V   & $3.88 \pm 0.10$ & $3.92 \pm 0.10$ & $6.19 \pm 0.11$       & $6.62 \pm 0.12$       & $-0.43\pm 0.16$       & early         &                         \\
          HAT-P-7               & F6V   & $3.88 \pm 0.10$ & $3.92 \pm 0.10$ & $11.53\pm 0.11$       & $10.79\pm 0.11$       & $0.74 \pm 0.16$       & M2-M5   & M5.5 [2]              \\
          HAT-P-8               & F5V   & $4.14 \pm 0.10$ & $4.50 \pm 0.10$ & $11.48 \pm 0.14$ & $11.18 \pm 0.12$   & $0.30 \pm 0.18$ [3]   & ---   &  M1-M3.5 [3,4]          \\
          HAT-P-41      & F9    & $4.35 \pm 0.10$ & $4.36 \pm 0.10$ & $8.00 \pm 0.20$       & $7.76 \pm 0.22$       & $0.25 \pm 0.30$       & F5-M2 &                         \\
          KIC 10905746  & K7    & $6.90 \pm 0.10$ & $6.58 \pm 0.10$ & $9.34 \pm 0.14$       & $8.88 \pm 0.14$       & $0.46 \pm 0.20$       & K5-M3 &                         \\
          TrES-2                & G0V   & $4.44 \pm 0.10$ & $4.44 \pm 0.10$ & $8.30 \pm 0.11$       & $7.80 \pm 0.11$       & $0.50 \pm 0.16$       & K7-M3   & M1-M2 [1], K0-M0 [3]  \\
          TrES-4                & F8    & $4.26 \pm 0.10$ & $4.28 \pm 0.10$ & $8.76 \pm 0.13$       & $8.44 \pm 0.13$       & $0.32 \pm 0.19$       & K2-M2   & M2 [1], K4.5-M1.5 [3] \\ 
          WASP-2                & K1.5  & $5.84 \pm 0.10$ & $5.71 \pm 0.10$ & $9.36 \pm 0.15$       & $8.97 \pm 0.19$       & $0.40 \pm 0.24$       & K2-M3   & M1V-M4V [3]    \\
          WASP-14               & F5V   & $3.69 \pm 0.10$ & $3.74 \pm 0.10$ & $9.03 \pm 0.11$       & $9.20 \pm 0.12$       & $-0.17\pm 0.16$       & early           &                       \\
          WASP-58               & G2V   & $4.54 \pm 0.10$ & $4.51 \pm 0.10$ & $10.20\pm 0.21$       & $9.59 \pm 0.16$       & $0.62 \pm 0.26$       & K7-M5   &                       \\
          \hline
                 
        \end{tabular}
      \begin{quote}
        *: Assuming physical companionship;
        [1]: \cite{Faedi13}, [2]: \cite{Narita12}, [3]: \cite{Bergfors13}, [4]: \cite{Bechter14}
      \end{quote}

      \end{center}
\end{table*}

\subsection{Probability of chance alignment}

For the newly discovered companion candidates, we only measured  one epoch  and can make no clear statement  on whether these stars are physically bound or unbound objects. We can, however, make a statistical argument on how many candidates of our whole sample are likely to be unrelated field stars. First, we estimate the density of field stars $\rho_i$ around each target by selecting all stars brighter than $m_K = 14\,$mag within $30'$ of the target in the 2MASS PSC \citep{Cutri03}. These criteria ensure that the stars are visible with AstraLux and are a good approximation of the mean stellar density around the source. We then assume that the stars are Poisson distributed and we calculate the probability of detecting one or more field stars $\overline{P_i(0)}$ (i.e.\ the complement of finding zero stars) with the maximum angular separation $\Theta$:
\begin{eqnarray}
    \overline{P_i(0)}=1-\exp(-\pi \Theta^2 \cdot \rho_i).
\end{eqnarray}
The probability of finding one or more field stars within our FoV of $6''$ ranges between $0.9 \%$ and $38 \%$; for the observed companion candidates with first epoch measurements, we find $38 \%$ for CoRoT-11, $14 \%$ for HAT-P-41, $4.4 \%$ for KIC 10905746, $0.9 \%$ for WASP-14, and $3.2 \%$ for WASP-38. The average probability of finding a field star is $4.7 \%$. For the whole survey of $42$ stars with first relative position measurement our expectation is thus two unrelated field stars. This means that out of our five objects one or two are likely not bound. Good candidates are CoRoT-11 and WASP-14 because of their early spectral type. In addition, the stellar density around CoRoT-11 is rather high. The seven follow-up targets with previous AstraLux epochs and previously known companion candidates (see Table \ref{no_comp}) were excluded from this analysis  to avoid  introducing any bias.

\subsection{System parameter changes of WASP-14 and WASP-58}

The newly discovered companion candidates will, if they are within the photometric aperture, contribute a constant flux offset that affects the stellar and planetary parameters derived from the depth and shape of the transit light curve: $R_p/R_*$, $b$, $a/R_*$ and $\rho_*$. The additional flux contribution from a stellar companion has the greatest effect on the planet-to-star radius; assuming no limb darkening and flat-bottomed transits, this parameter is directly proportional to the depth of the transit \citep[see][]{Seager03},\\

\begin{equation}
\Delta F = (F_{no \, \, transit}-F_{transit})/F_{no \, \, transit} = (R_p/R_*)^2
.\end{equation}

We account for the secondary star flux using the relation between $\Delta F_{old}$, assuming only the flux from the brighter planet host, and $\Delta F_{new}$, including the companion flux \citep[see][]{Daemgen09},\\

\begin{equation}
\Delta F_{new} = (1+10^{-\Delta z/2.5})*\Delta F_{old}
,\end{equation}

where $\Delta z$ is our measured magnitude difference between the planet host star and the companion candidate in $z$-band (see Table~\ref{with_comp}). We calculate new planet-to-star ratios for the WASP-14 and WASP-58 systems using $\Delta F$ from the planet discovery papers by \citet[WASP-14b;][]{Joshi09} and \citet[WASP-58b; ][]{Hebrard13}. The new companions are both $>5$\,mag fainter than the planet host stars and therefore contribute comparatively little flux. For the WASP-14 system we find that, after accounting for the previously unknown companion candidate,   the planet-to-star ratio $R_*/R_p$  increases by 0.3\%, which is well within the $\approx1.4\%$ errors derived by \cite{Joshi09}. Similarly, the companion candidate to WASP-58 only adds 0.5\% to the same parameter, therefore not affecting the results by \cite{Hebrard13}.

\section{Discussion and summary}

A stellar companion to an exoplanet hosting system is expected to affect planet formation and the subsequent dynamical evolution in several ways. Our observations of transiting planet host stars are part of an ongoing effort to find and characterise binary companions using the Lucky Imaging technique. During this run we took high-resolution images of $49$ exoplanet host stars, which revealed two previously unknown stellar companion candidates to the sources WASP-14 and WASP-58. Close, undetected sources, whether bound or not, influence the derivation of stellar and planetary properties of the host system. For WASP-14 and WASP-58 we show, however, that their presence only has a minor impact because of their faintness. Apart from these two sources we also measured the astrometric position of CoRoT-11, HAT-P-41, and KIC 10905746 for the first time. We deduce from the mean stellar density around our targets that three out of those five candidate companions are likely to be gravitationally bound. This is also supported by the fact that HAT-P-41, KIC-10905746, and WASP-58 have a spectral type of late K or early M as expected for a fainter associated companion. By contrast, CoRoT-11 is situated in a dense stellar region and the detected source at $2.5\arcsec$ most likely has an early spectral type; it is thus unlikely to be bound. 

In addition, we did follow-up observations of CoRoT-2, CoRoT-3, HAT-P-7, HAT-P-8, TrES-2, TrES-4, and WASP-2. Our spectral type estimations for these sources agree with earlier observations. The new positional data for CoRoT-2, CoRoT-3, and TrES-2 is not sufficient to decide on common proper motion, which is already known for HAT-P-7 \citep{Narita12}, HAT-P-8 \citep{Ginski13, Bechter14}, TrES-4 \citep{Bergfors13}, and WASP-2 \citep{Bergfors13}

The knowledge of gravitationally bound companions can be used to correlate binarity and exoplanet properties in future. This analysis is, however, hampered by the fact that no orbital motion of the stellar companions is visible so far. Thus, for now the true stellar separation and the relative orientation of the planetary and the stellar orbit remain unknown.

\bibliographystyle{aa}
\bibliography{lit}

\end{document}